\documentclass[twocolumn,showpacs,preprintnumbers]{revtex4}
\usepackage{amssymb}
\usepackage{amsmath}
\usepackage{graphicx}
\usepackage{dcolumn}
\usepackage{bm}

\begin{document}

\preprint{APS/123-QED}
\title{Quantum point contact conductance in NINS junctions}
\author{William J. Herrera}
\email{jherreraw@unal.edu.co}
\author{J. Virgilio Ni\~{n}o}
\email{jvninoc@unal.edu.co}
\author{J. Jairo Giraldo}
\email{jjgiraldog@unal.edu.co}
\affiliation{Departamento de F\'{\i}sica, Universidad Nacional de Colombia, Bogot\'{a}-
Colombia.\\
}
\date{\today }

\begin{abstract}
The effect of an insulating barrier located at a distance $a$ from a NS
quantum point contact is analyzed in this work. The Bogoliubov de Gennes
equations are solved for NINS junctions (S: anysotropic superconductor, I:
insulator and N: normal metal), where the NIN region is a quantum wire. For $%
a\neq0$, bound states and resonances in the differential conductance are
predicted. These resonances depend on the symmetry of the pair potential,
the strength of the insulating barrier and $a $. Our results show that in a
NINS quantum point contact the number of resonances vary with the symmetry
of the order parameter. This is to be contrasted with the results for the
NINS junction, in which only the position of the resonances changes with the
symmetry.
\end{abstract}

\pacs{74.20.Rp,74.50.+r,74.45.+c,81.07.Lk}
\maketitle




\section{\label{sec:int}Introduction}

In high critical temperature superconductivity the symmetry of the pair
potential is one the most studied aspects \cite{Tsuei,Van}. Tunneling spectra depend strongly on this symmetry and therefore tunneling spectroscopy is a very sensitive probe for its
study. In a $d$-symmetry and (110) orientation, for instance, the
differential conductance has a peak at zero voltage, called zero-bias
conductance peak (ZBCP) which has been predicted theoretically by differents works \cite%
{Tanaka1,Kashiwaya1,Kashiwaya2,Kashiwaya3,Barash,Walker} and observed
experimentally \cite{Covin,Alff,Wei,Aprili,Wang,Iguchi} . The existence of
the ZBCP is due to the formation of Andreev bound states at the Fermi level
(zero energy states) near to the interface \cite{Yang, Hu, Tanaka2}. These
states appear due to the interference between scattering quasiparticles at
the interface and the sign change of the pair potential. Studies of quantum
point contacts in NIS junctions show that the ZBCP is removed by the
quasiparticle diffractions at the point contact \cite{Takagaki,Tsuchikawa},
an aspect that has been shown experimentally \cite{Iguchi}. Recently two
quantum point contacts have been studied for the crossed Andreev reflection
in $d$-wave superconductors \cite{Takahashi}.

On the other hand, in NINS \cite{Ridel,Tessmer} and NISN junctions \cite%
{McMillan,Herrera}, resonances in the differential conductance appear. In
anisotropic superconductors, the resonance energies depend as well on the
symmetry of the pair potential. In NINS junctions and $d_{xy}$-symmetry,
e.g. , the positions of these resonances are out of phase with respect to
those predicted for isotropic superconductors\cite{Xu} and in NISN junctions
the conductance presents two kinds of resonances due to anisotropy of the
pair potential \cite{Herrera}.

In this paper, we analyze the differential conductance when quasiparticles
are injected into a superconductor from a single-mode quantum wire, with an
insulating barrier located at a distance $a$ of the NS interface (NINS
quantum point contact). We show that there exist bound states which cause
resonances in the differential conductance and that the number of these
resonances depends on the symmetry of the order parameter. This is shown through 
the solution of the Bogoliubov-de Gennes equation in NINS
junctions, where NIN region is modeled by a wire of width $W$. In particular 
$s$ and $d$ - symmetries are considered.

\section{\label{sec:BdGE}The Bogoliubov-de Gennes equation and its solutions
in NINS point contacts}

The elementary excitations or quasiparticles in a superconductor are
described by the Bogoliubov de Gennes (BdG) equations , which can be
generalized for anisotropic superconductors \cite{Bruder}. For a steady
state these equations are

\begin{equation}
\begin{split}
H_{e}(\mathbf{r}_{1})u(\mathbf{r}_{1})+\int d\mathbf{r}_{2}\Delta (\mathbf{r}%
_{1},\mathbf{r}_{2})v(\mathbf{r}_{2})& =Eu(\mathbf{r}_{1})\ , \\
-H_{e}^{\ast }(\mathbf{r}_{1})v(\mathbf{r}_{1})+\int d\mathbf{r}_{2}\Delta (%
\mathbf{r}_{1},\mathbf{r}_{2})u(\mathbf{r}_{2})& =Ev(\mathbf{r}_{1})\ ,
\end{split}%
\end{equation}%
where $H_{e}=-\hbar ^{2}\nabla ^{2}/2m+V\left( \mathbf{r}\right) -\mu $ is
an electronic hamiltonian and $\mu $ the chemical potential. $\Delta (%
\mathbf{r}_{1},\mathbf{r}_{2})$ is the pair potential, $u(\mathbf{r}_{1})$
and $v(\mathbf{r}_{1})$ are the wave function for the electron- and
hole-like components of a quasiparticle,

\begin{equation}
\psi (\mathbf{r})=%
\begin{pmatrix}
u(\mathbf{r}) \\ 
v(\mathbf{r})%
\end{pmatrix}%
\text{\ .}
\end{equation}

\begin{figure}[tbh]
\begin{center}
\includegraphics{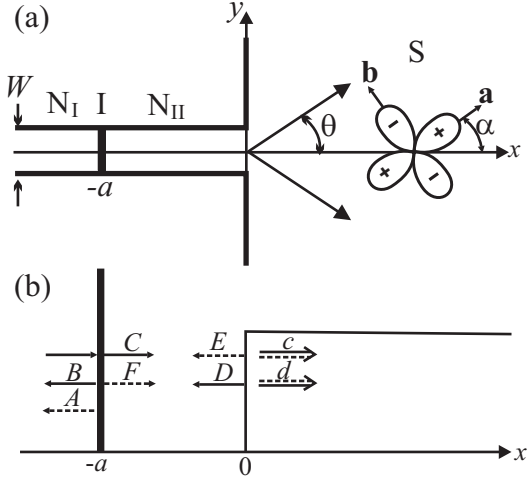}
\end{center}
\caption{\textit{(a) The point contact NINS junction, the insulating barrier
is located in $x=-a$ and the NIN region is a single mode quantum wire with
width $W$. (b) Scattering processes, the solid line and dashed line
represent the electron and the hole-like components of a quasiparticle
respectively.}}
\label{fig:Junction}
\end{figure}

It is suppossed that the quasiparticle moves on the $x$-$y$ plane, the
interfaces are normal to the $x$-axis and the NIN region has a width $W$ in
the $y$ direction, see fig. \ref{fig:Junction}. The insulating barrier is
modeled by a delta function, $V(x)=U_{0}\delta (x+a)$. The solutions of the
BdG equations in the N$_{I}$, N$_{II}$ and in the superconducting regions,
are respectively,

\begin{equation}
\psi _{N_{I}}=\left[ 
\begin{pmatrix}
1 \\ 
0%
\end{pmatrix}%
e^{ik_{1}^{+}x}+A%
\begin{pmatrix}
0 \\ 
1%
\end{pmatrix}%
e^{ik_{1}^{-}x}+B%
\begin{pmatrix}
1 \\ 
0%
\end{pmatrix}%
e^{-ik_{1}^{+}x}\right] \phi _{1}(y)  \label{pN1}
\end{equation}

\begin{align}  \label{pN2}
\psi _{N_{II}}=&\left[ C 
\begin{pmatrix}
1 \\ 
0%
\end{pmatrix}%
e^{ik_{1}^{+}x}+D%
\begin{pmatrix}
1 \\ 
0%
\end{pmatrix}%
e^{-ik_{1}^{+}x}+E%
\begin{pmatrix}
0 \\ 
1%
\end{pmatrix}%
e^{ik_{1}^{-}x}\right.  \notag \\
&+ \left. F%
\begin{pmatrix}
0 \\ 
1%
\end{pmatrix}%
e^{-ik_{1}^{-}x}\right] \phi _{1}(y),
\end{align}

\begin{align}  \label{psiS}
\psi _{S}=&\int_{-k_{F}}^{k_{F}}ds\left[ c(s)%
\begin{pmatrix}
u_{0}^{+}(s)e^{i\varphi _{+}(s)/2} \\ 
v_{0}^{+}(s)e^{-i\varphi _{+}(s)/2}%
\end{pmatrix}%
e^{ik_{+}^{+}(s)x}+\right.  \notag \\
&d(s)\left. 
\begin{pmatrix}
v_{0}^{-}(s)e^{i\varphi _{-}(s)/2} \\ 
u_{0}^{-}(s)e^{-i\varphi _{-}(s)/2}%
\end{pmatrix}%
e^{-ik_{-}^{-}(s)x}\right] e^{isy},
\end{align}

where 
\begin{align}  \label{kas}
& k_{1}^{\pm }=\sqrt{k_{1}^{2}\pm 2mE/\hbar ^{2}},\; k_{1}=\sqrt{%
k_{F}^{2}-\pi ^{2}/W^{2}},  \notag \\
& k_{\pm }^{\pm }(s)=\sqrt{k_{1}^{2}\pm 2m\Omega _{\pm }(s)/\hbar ^{2}}.
\end{align}
\begin{align}
&\phi _{1}(y)=\sqrt{\frac{2}{W}}\sin \left[ \frac{\pi }{W}\left( y+\frac{W}{2%
}\right) \right] , \\
&\Omega _{\pm }(s)=\sqrt{E^{2}-\left\vert \Delta _{\pm }(s)\right\vert ^{2}},
\end{align}

\begin{equation}
u_{0}^{\pm }(s)=\sqrt{\frac{1}{2}\left[ 1+\frac{\Omega _{\pm }(s)}{E}\right] 
},~~v_{0}^{\pm }(s)=\sqrt{\frac{1}{2}\left[ 1-\frac{\Omega _{\pm }(s)}{E}%
\right]}~~.
\end{equation}

The quasiparticles with $k_{+}^{+}$ and $k_{-}^{-}$wavenumber move in the
pair potential $\Delta _{+}$ \ and $\Delta _{-}~\ $respectively and given by

\begin{align}
\Delta _{\pm }(s)=\Delta (\pm k_{\pm}^{\pm}\mathbf{\hat{\imath}}+s\mathbf{%
\hat{\jmath}})\equiv\Delta _{\pm }e^{i\varphi _{\pm }},  \notag \\
\intertext{ with }
\Delta (\mathbf{k})=\int d(\mathbf{r}_{1}-\mathbf{r}_{2})e^{i\mathbf{k}\cdot
(\mathbf{r}_{1}-\mathbf{r}_{2})}\Delta (\mathbf{r}_{1}-\mathbf{r}_{2})\text{%
\ .}
\end{align}

All the evanescent modes have been neglected. This approximation is
justified due to the fact that for $\pi <Wk_{F}<2\pi $ \ \ the narrow wire
is a single mode and the energy of evanescent modes is well above the Fermi
energy \cite{Takagaki,Takahashi}. One finds $A$, $B$, $C$, $D$, $E$, $F$, $c$
and $d$ using boundary condictions in $x=-a$ and $x=0$. The
electron-electron and electron-hole reflection coefficients are respectively

\begin{equation}
R_{e}=\left\vert \frac{h}{g}\right\vert ^{2},\ R_{h}=\left\vert \frac{2F_{3}%
}{g}\right\vert ^{2}\text{,}
\end{equation}

where

\begin{align}  \label{g}
g =&(1+Z^{2})\left[(1+F_{1})^{2}-F_{2}F_{3}\right]  \notag \\
&+Z^{2}\left[(1-F_{1})^{2}-F_{2}F_{3}\right]e^{-2i(k_{+}-k_{-})a}  \notag \\
&+Z\left[1-F_{1}^{2}+F_{2}F_{3}\right]\left[Z(e^{2ik_{-}a}+e^{-2ik_{+}a})%
\right.  \notag \\
&\left.+i(e^{-2ik_{+}a}-e^{2ik_{-}a})\right] \text{,}
\end{align}

\begin{align}  \label{h}
h =&(F_{1}^{2}-F_{2}F_{3}-1)[Z^{2}e^{2i(k_{+}+k_{-})a}-(1-iZ)^{2}]  \notag \\
&-Z(Z+i)\left[2F_{1}(e^{2ik_{+}a}-e^{2ik_{-}a})\right.  \notag \\
&\left.+(1+F_{1}^{2}-F_{2}F_{3})(e^{2ik_{+}a}+e^{2ik_{-}a})\right] \text{,}
\end{align}

\begin{equation}
F_{i}=\frac{4}{\pi ^{2}\sqrt{\gamma _{F}^{2}-1}}\int_{-\gamma _{F}}^{\gamma
_{F}}dq\frac{\sqrt{\gamma _{F}^{2}-q^{2}}}{(1-q^{2})^{2}}\cos ^{2}[\pi
q/2]f_{i}(q)\text{, \ }  \label{Fi}
\end{equation}

\begin{align}
f_{1}=&\frac{1+\Gamma _{+}\Gamma _{-}e^{-i(\varphi _{+}-\varphi _{-})}}{%
1-\Gamma _{+}\Gamma _{-}e^{-i(\varphi _{+}-\varphi _{-})}}\text{, \ }f_{2}=%
\frac{2\Gamma _{-}e^{i\varphi _{-}}}{1-\Gamma _{+}\Gamma _{-}e^{-i(\varphi
_{+}-\varphi _{-})}} \text{,}  \notag \\
f_{3}=&\frac{2\Gamma _{+}e^{-i\varphi _{+}}}{1-\Gamma _{+}\Gamma
_{-}e^{-i(\varphi _{+}-\varphi _{-})}} \text{,}
\end{align}

\begin{equation}
\Gamma _{\pm }=\frac{v_{0}^{\pm }}{u_{0}^{\pm }}\text{, }\gamma _{F}=\frac{%
k_{F}W}{\pi }\text{\ \ and \ }Z=\frac{mU_{0}\gamma _{F}}{\hbar ^{2}m\sqrt{%
\gamma _{F}^{2}-1}}.
\end{equation}

\section{\label{sec:DC}Diferential conductance}

Using the BTK model \cite{BTK}, the normalized differential conductance, $%
G_{R},$ is calculated from 
\begin{equation}
G_{R}=\frac{G_{S}}{G_{N}}=\frac{[(1+F_{0})^{2}+4Z^{2}]\left(
1-R_{e}+R_{h}\right) }{4F_{0}},
\end{equation}%
where $G_{N}$ \ the conductance when $\Delta =0$ and \ $a=0$. $F_{0}$ is
defined by (\ref{Fi}) with $f_{i}=1$. For $d$-symmetry $\Delta _{\pm
}=\Delta _{0}\cos (2(\theta \mp \alpha ))$, $\alpha $ is the angle between
the (100) axis of the superconductor and the normal to the interface, and $%
\theta =\sin ^{-1}(s/k_{F})$. (cf. fig. \ref{fig:Junction}a.)

\begin{figure}[tbh]
\begin{center}
\includegraphics{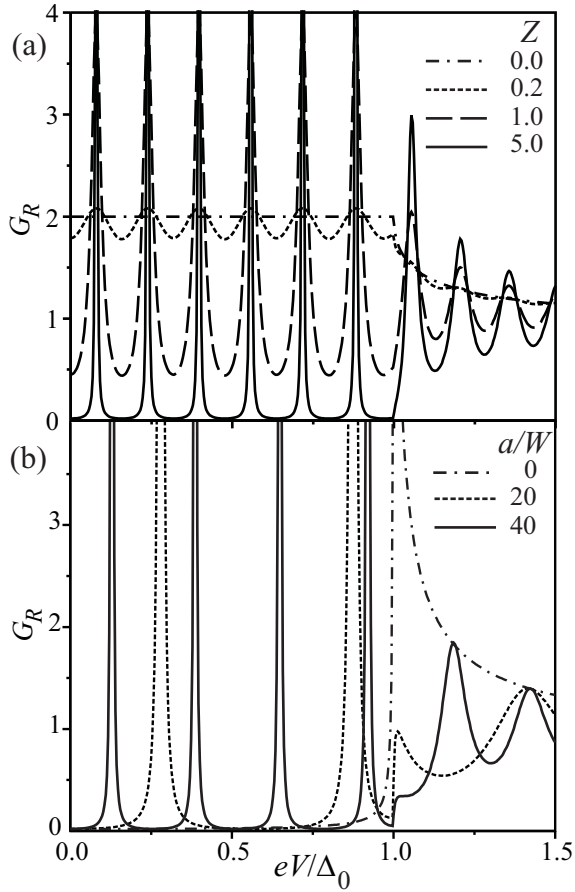}
\end{center}
\caption{\textit{Diferential conductance for $s$-symmetry. (a) Different
values of $Z$ with $a=63W$ ; (b) different values of $a$ with $Z=5$. In both
cases $k_{F}W=1.7$.}}
\label{fig:DCs}
\end{figure}

\begin{figure}[htb]
\begin{center}
\includegraphics{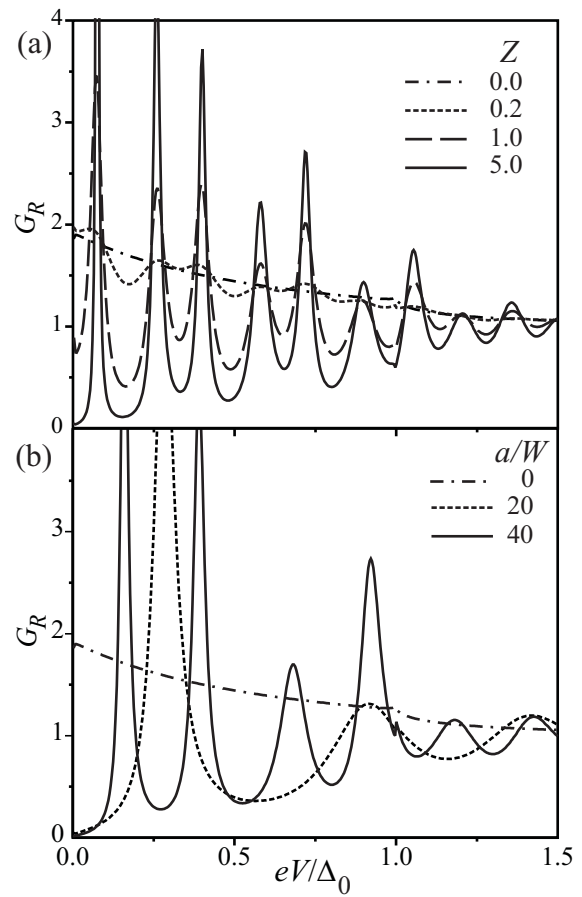}
\end{center}
\caption{\textit{Same as in fig. \protect\ref{fig:DCs} for $\protect\alpha=0 
$ ($d_{x^{2}-y^{2}}$-symmetry).}}
\label{fig:DCd0}
\end{figure}

\begin{figure}[htb]
\begin{center}
\includegraphics{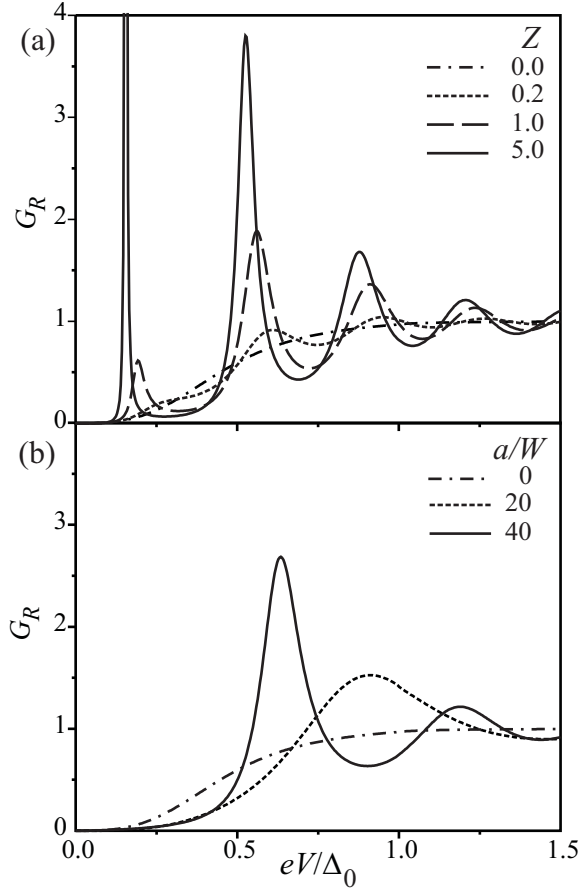}
\end{center}
\caption{\textit{Same as in fig. \protect\ref{fig:DCs} for $\protect\alpha=%
\protect\pi/4$ ($d_{x-y}$-symmetry).}}
\label{fig:DCdpi2}
\end{figure}

\begin{figure}[hbt]
\begin{center}
\includegraphics{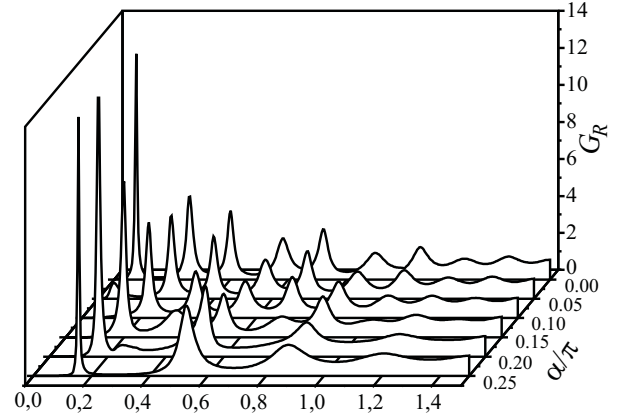}
\end{center}
\caption{\textit{Diferential conductance for different values of $\protect%
\alpha$ with $Z=5$, $a=63W$ and $k_{F}W=1.7$.}}
\label{fig:DC3D}
\end{figure}

Figures \ref{fig:DCs} and \ref{fig:DCd0} show the differential conductance
for $s$ and $d_{x^{2}-y^{2}}$-symmetries. When $a=0$ (NIS point contact),
our results agree with \cite{Takagaki}. For $a\neq 0$, subgap resonances
appear in the differential conductance and their number increases with $a$.
When $Z$ decreases, it can be seen that the number and position of the
resonances remain unchanged, only the peaks become broaden and for $%
d_{x^{2}-y^{2}}$-symmetry the broad is greater for a fixed $Z$ value.

Figure \ref{fig:DCdpi2} exhibits the differential conductance for $\alpha
=\pi /4$ $(d_{x-y}$-symmetry). ZBCP\ does no apppear because its Andreev
reflection is zero. In the last case the wave functions in the channel are a
superposition of two plane waves with wave numbers $k_{y}=\pm \pi /W$, each
wave experiences \ a pair potential phase $0$ and $\pi $ respectivley and
therefore the Andreev reflection coefficient $a(\theta )$ ($R_{h}(\theta
)=\left\vert a(\theta )\right\vert ^{2}$) for each wave are opposite, the
waves of the reflected holes interfere destructively and the Andreev
reflections vanish. In relation with the $d_{x-y}$ - symmetry the number of
resonances decreases compared \ whit the $s$ and $d_{x^{2}-y^{2}}-$%
symmetries. Additionally when $Z$ decreases the number of resonances is
constant and the peak broadens and its position is smoothly displaced toward
the right.

The subgap resonances in the differential conductance are a direct
consequence of the quasibound states formed inside the energy gap. The
energies and lifetime of these bound states are given by the poles of the
current transmission amplitude. Setting $g=0$ in Eq.\ref{g} one finds these
poles. A complex energy, $E=E_{R}+iE_{I}$, is required in order to solve
this equation, where $E_{R}$ is the position of resonance and $\hbar
/2\left\vert E_{I}\right\vert $ is the lifetime of the quasibond states.

The resonance positions $E_{R}$ for $s$ or $d_{x^{2}-y^{2}}$-symmetries are
given by

\begin{equation}
E_{n}=E_{0}(n\pi -\phi ),n=1,2,...,
\end{equation}%
and for $d_{xy}$ symmetry are determined from 
\begin{equation}
E_{n}=E_{0}(2n\pi -\phi ^{\prime }),\ n=1,2,...
\end{equation}%
In these equations 
\begin{equation}
\ E_{0}=\frac{E_{F}}{a}\sqrt{1-\gamma _{F}^{-2}}\,
\end{equation}%
and $\phi $ , $\phi ^{\prime }$ are phases that depend on $Z$, $a$ and $E$.
Therefore the number of resonances with $E<max(\Delta )$ for $%
s$ or $d_{x^{2}-y^{2}}$-symmetries are approximately twice the corresponding
number of a $d_{xy}$-symmetry. This is due to the fact that in the case of a 
$d_{xy}$-symmetry the Andreev reflection is zero. Thereby the quantization
of the bound states occurs when the quasiparticles travel in a closed path a
distance equal to $2a$ in the $x$ direction and $E_{n}\propto 2n\pi /2a$.
(The quasiparticle is transmitted as an electron in $x=-a$ and reflected as
an electron in $\ x=0$.) In the case of $\ s$-or $d_{x^{2}-y^{2}}$-
symmetries the quasiparticles complete a closed path when they travel a
distance $\ 4a$ in $x$ direction and $E_{n}\propto 2n\pi /4a$. \ (The
quasiparticle is transmitted in $x=0$ as an electron, reflected as a hole in 
$x=\ 0$, reflected as a hole in $\ x=-a$ and finally reflected as an
electron in $x=0$.) Therefore one has that in this case the number of the
quasibound states is approximately twice the corresponding number of $d_{x-y}
$ symmetry.

\bigskip

In order to determine the lifetime of the quasibound states, a semiclassical
argumentation will be used. The lifetime $\tau$ is defined as the time that
a quasiparticle in the INS region requires to "scape" toward the $N_{I}$ or $%
S$ regions. For the $s$ or $d_{x^{2}-y^{2}}$- symmetries the time that a
quasiparticle needs for around trip is

\begin{equation}
T=\frac{4a}{\hslash k_{0F1}/m}=\frac{2\hslash d}{E_{F}\sqrt{1-\gamma
_{F}^{-2}}}.  \label{eq:Ts}
\end{equation}

If $N$ is the number of closed trips,  $\tau $ is given by%
\begin{equation}
\tau =TN,  \label{eq:taudef}
\end{equation}
$N$\ is obtained from 
\begin{equation}
\left[ R_{e-h}R_{h-e}R_{h-h}R_{e-e}\right] ^{N}=1/e,  \label{eq:N}
\end{equation}%
where $R_{e-h},\;R_{h-e}$ are the electron-hole and hole-electron reflection
coefficients respectively for $Z=0$ (point contact NS) and $%
RI_{e-e},\;RI_{h-h}$ the electron-electron and hole-hole reflection
coefficients respectively for an insulating barrier (IN). From equations (%
\ref{eq:Ts}), (\ref{eq:taudef}) and (\ref{eq:N}) the lifetime is obtained as

\begin{equation}
\tau =-\frac{\hslash d}{E_{F}\sqrt{1-\gamma _{F}^{-2}}}\frac{1}{\ln \left(
Z^{2}R_{e-h}/\left( 1+Z^{2}\right) \right) },  \label{eq:taus}
\end{equation}%
where we have used the fact that $R_{e-h}=R_{h-e}$ and $%
RI_{e-e}=RI_{h-h}=Z^{2}/\left( 1+Z^{2}\right) $. Equation (\ref{eq:taus}) is \ 
similar to that found for NINS junction with $s$-symmetry \cite{Ridel}.
Similarly for the $d_{xy}$-symmetry the lifetime is given by

\begin{equation}
\tau =-\frac{\hslash d}{E_{F}\sqrt{1-\gamma _{F}^{-2}}}\frac{1}{\ln \left(
Z^{2}R_{e-e}/\left( 1+Z^{2}\right) \right) },
\end{equation}%
with $R_{e-e}$ the electron-electron reflection coefficient for $Z=0$. For
the case of $s-$ symmetry $,$ and $E<\left\vert \Delta \right\vert ,$ $\
R_{e-h}=1$. Therefore the lifetime increases with $Z$ and tends to infinity
for $Z>>1$, while the resonance width, $2\left\vert E_{I}\right\vert \approx
\hbar /\tau \rightarrow 0$, as is observed in fig. \ref{fig:DCs}. For $%
d_{x^{2}-y^{2}}-$ symmetry the quasiparticles transmission is finite for $%
E<\Delta _{0}$ due to the anisotropy of the pair potential,  $R_{e-h}<1$,
and the lifetime increases with $Z$ but is finite for $Z>>1$. This is
observed in the width of the resonances in fig. \ref{fig:DCd0}. For the $%
d_{xy}-$ symmetry the behavior of the lifetime and the width of the
resonances are similar to the case of $d_{x^{2}-y^{2}}$- symmetry, see Fig. %
\ref{fig:DCdpi2}. For all cases, with $E>\Delta _{0},$ the reflection
coefficients are always less that one, the lifetimes decrease and the widths
of the resonances increase.

Figure \ref{fig:DC3D} shows $G_{R}$ for different values of $\alpha $, when $%
\alpha $ change from $\ 0$ \ to $\pi /4$ ,\ some peaks begin to decrease and
vanish for $\alpha \approx 0.20\pi $. This happens because the Andreev
reflections decrease and the electron-electron reflection increases. For $%
\alpha =\pi /4$ the Andreev reflections are zero and one has the conductance
for $d_{xy}$- symmetry. Similarly the values of the energy of the resonances
move toward the left as $\alpha $ increases due to a change of the phase $%
\phi $ in the solution of the equation $g=0$.

\section{Conclusions}

Our results show that in NINS point contacts the differential conductance
have resonances due to bound states. The number of resonances depends on the
symmetry of the order parameter, in contrast to a NINS junction. In the
latter case only the position of the resonances changes with the symmetry.
The number of resonances with $E<max(\Delta )$ (subgap resonances) for $s$
or $d_{x^{2}-y^{2}}$-symmetries is approximately twice the corresponding
number of the $d_{xy}$-symmetry. When $\alpha $ change from$\ 0$ to $\pi /2$
some peaks dissapear because to Andreev reflection vanishes

In the case of $s-$symmetry, the lifetime of quasibound states increases with
the insulating barrier strength and is infinite for $Z>>1$. In contrast, for
a $d$-symmetry the lifetime increases with $Z$ but is finite for $Z>>1$.
This occurs because the quasiparticles transmission is different of zero for 
$E<\Delta _{0}$ in contrast to the case of  $s$-symmetry, where the
transmission is zero for $E<\Delta _{0}.$ Therefore the lifetime of the
resonances decreases in $d$-symmetries and the width of the resonances
increases. The results obtained in this work can be used to find the
symmetry of high temperature superconductors in experiments of the type
carried out in references \cite{Alff} and \cite{Wei}.

\begin{acknowledgments}
We are grateful to Divisi\'{o}n de investigaciones de la sede Bogot\'{a} de
la  Universidad Nacional de Colombia for supporting this work. 
\bibliographystyle{apsrev}
\bibliography{Pointrevtex}
\end{acknowledgments}

\end{document}